# Molecular Dynamics Simulation Study on the Melting of Ultra-thin Copper Nanowires

Jeong Won Kang* and Ho Jung Hwang

Semiconductor Process and Device Laboratory, Department of Electronic Engineering,

Chung-Ang University, 221 HukSuk-Dong, DongJak-Ku, Seoul 156-756, Korea

## ABSTRACT

We have investigated the melting behavior of ultra-thin copper nanowires using classical molecular dynamics simulations. The caloric curves of cylindrical multi-shell copper nanowires showed an insight into the specific phase transition. The melting temperature of copper nanowires is linearly proportional with the number of atoms per layer. When nanowires have almost the same number of atoms per layer regardless of the initial structures, the melting temperatures of nanowires are much the same.

In studies of nanoelectronic-device fabrications, metallic nanowires (NWs) have been extensively investigated over the past decade [1-25]. The structure of ultra-thin metallic NWs has been investigated using molecular dynamics (MD) simulations [2-15,20-25]. Previous works have shown that ultra-thin Cu NWs obtained from atomistic simulations under various conditions can have several structures such as face-centred-cubic (*fcc*) structure [24], pentagonal multi-shell (PMS)-type [24,25], and cylindrical multi-shell (CMS)-type [22-25], etc.

Melting behavior of NWs have been investigated using the MD simulations for NWs of Au [9] and Pb [15]. All nanowires have molten below the bulk melting temperature [9,15]. Copper is an important engineering material, so in this investigation we present a specific phase transition and the melting behavior of ultra-thin Cu NWs using MD simulations.

We have performed several MD runs at different temperatures and for different radii and structures of wires. We used twelve NWs with fcc, PMS-type, and CMS-type, as shown in Table 1. This MD simulation used the same MD methods as our previous works [20-25,27-32] with time step of $10^{-15}$ s. The periodic boundary condition (*pbc*) was applied along the axes of NWs. The lengths of supercells for Cu NWs with {100}, {110}, and {111} are 36.150, 61.343, and 41.738 Å, respectively. The lengths of supercells of Cu NWs with PMS-type and CMS-type structures are 63.284 and 66.512 Å, respectively. The initial structures were first relaxed by the MD steps of 200 000 at 300 K using NWs with the ideal atomic positions. Each system was then heated by the MD runs of 10 000 time steps (10 ps) with a temperature step of 1 K from 300 K. We monitored the internal energy as a function of temperature. Interaction between Cu atoms was described by a well-fitted potential function of the second moment approximation of the tight-binding (SMA-TB) scheme [26], which has been used in our previous works [20-25, 27-32].

Figure 1 shows the caloric curves for some of ultra-thin Cu NWs in Table 1. In the caloric curves, we can identify four regions, labeled as *A*, *B*, *C*, and *D* in Fig. 1. In region *A*, the ultra-thin nanowire is solid, and the slope corresponds to the Gulong-Petit specific heat. In the region *B*, the caloric curve exhibits an upward curvature where the specific heat pronouncedly increases by the beginning of melting. In the region *C*, the ultra-thin nanowire is melting state, and the slope is the same with that in the region *A*. In the region *D*, the ultra-thin nanowire is broken and then a spherical cluster is formed. Since the *pbc* were applied to the supercells for the MD simulations, a spherical cluster in the region *D* is shown. The slope in the region *D* is the same with that in the region *A* and *C*. When the number of atoms per layer ($N$) for NWs is below 16, regions *B* and *C* are not shown. Since $N$ of NW is very small, as soon as temperature reaches the melting point of NW, NW is broken and then spherical cluster is formed. In NWs with $N$ above 18, all specific regions, such as *A*, *B*, *C*, and *D* regions, appear. However, in the cases including some structural transformations, such as *fcc* to CMS-type and PMS-type to CMS-type, the slope of caloric curve is slightly greater than the normal slope before the region *B*. These phenomena are shown in ultra-thin Cu NWs with initially FCC and PMS-type structures. All transformed NWs don't have CMS-type structure. However, most of them have the structures mixed by *fcc* or CMS-type. Cu NWs with the mixed-structures were shown in previous work [24].

Figure 2 shows the snapshots of $C_{34}$ NW for temperatures and would be mish helpful to understand the melting behavior of NWs. Figure 2(a) shows that the structure of the solid region, *A*, at 450 K maintains the initial structure. However, at 580 K, the melting temperature of $C_{34}$ NW, the structure is deformed as shown in Fig. 2(b). At 700 K, $C_{34}$ NW is in the melting state, the structure is similar to an amorphous state as shown in Fig. 2(c). Figure 2(d) shows the structure just before the breaking of NW at 800 K. As the temperature reaches near the breaking point of NW, the neck of NW becomes more and more narrow and finally is broken as the fuse by the heat burns out. As mentioned above, Fig. 2(e) shows a spherical cluster formed after the breaking of NW. The spherical cluster is induced by the *pbc* applied to the supercell.

Figure 2 shows the melting temperature ($T_{Mnw}$) of NWs as a function of the number of atoms per layer ($N$). In Fig. 2, $T_{Mnw}$ is linearly proportional with $N$, $T_{Mnw}= 4.5\ N + 420$. When NWs have almost the same $N$ regardless of the initial structure, the melting temperatures of NWs are much the same in height. This implies that $T_{Mnw}$ is related to more the initial diameter than the initial structure.

This brief report showed the specific states achieved by the temperature elevation from the caloric curves. The melting temperatures of ultra-thin Cu nanowires are much lower than the bulk melting temperature, and are linearly proportional with the number of atoms per layer. When the number of atoms per layer of nanowires is below 16, a nanowire is broken as soon as the temperature reaches the melting point of the nanowire. In cases including the structural transformations of nanowires, the slope of caloric curves of the nanowire is slightly greater than normal slope. The caloric curves of cylindrical multi-shell Cu nanowires with above four shells give an insight into a specific phase transition, solid to melting. From the result that cylindrical multi-shell Cu nanowires maintain their structures at below the melting temperature, we can expect to apply to Cu nanowires to nanoelectronic devices.

**Table**

Table I. Several ultra-thin Cu nanowires simulated in the work.

|  | Structure | Atoms/layer | Cross-section shape |
|---|---|---|---|
| $N_{18\{100\}}$ | {100} | 18 | Octagon |
| $N_{50\{100\}}$ | {100} | 50 | Octagon |
| $N_{18\{110\}}$ | {110} | 18 | Hexagon |
| $N_{54\{110\}}$ | {110} | 54 | Hexagon |
| $N_{13\{111\}}$ | {111} | 13 | Hexagon |
| $N_{49\{111\}}$ | {111} | 49 | Hexagon |
| $P_6$ | PMS-type 5-1 | 6 | Pentagon |
| $P_{16}$ | PMS-type 10-5-1 | 16 | Pentagon |
| $P_{51}$ | PMS-type 20-15-10-5-1 | 51 | Pentagon |
| $C_7$ | CMS-type 6-1 | 7 | Circle |
| $C_{18}$ | CMS-type 11-6-1 | 18 | Circle |
| $C_{34}$ | CMS-type 16-11-6-1 | 34 | Circle |

Figure Captions

Figure 1. The caloic curves, i. e. energy per atom as a function of temperature for some nanowires in Table. I

Figure 2. Snapshots of $C_{34}$ nanowire for the temperature. (a) 450 K. (b) 580 K. (c) 700 K. (d) 800 K. (e) 850 K.

Figure 3. The melting temperature of naowires ($T_{Mnw}$) in Table. I as a function of the number of atom per layer ($N$).

FIGURES

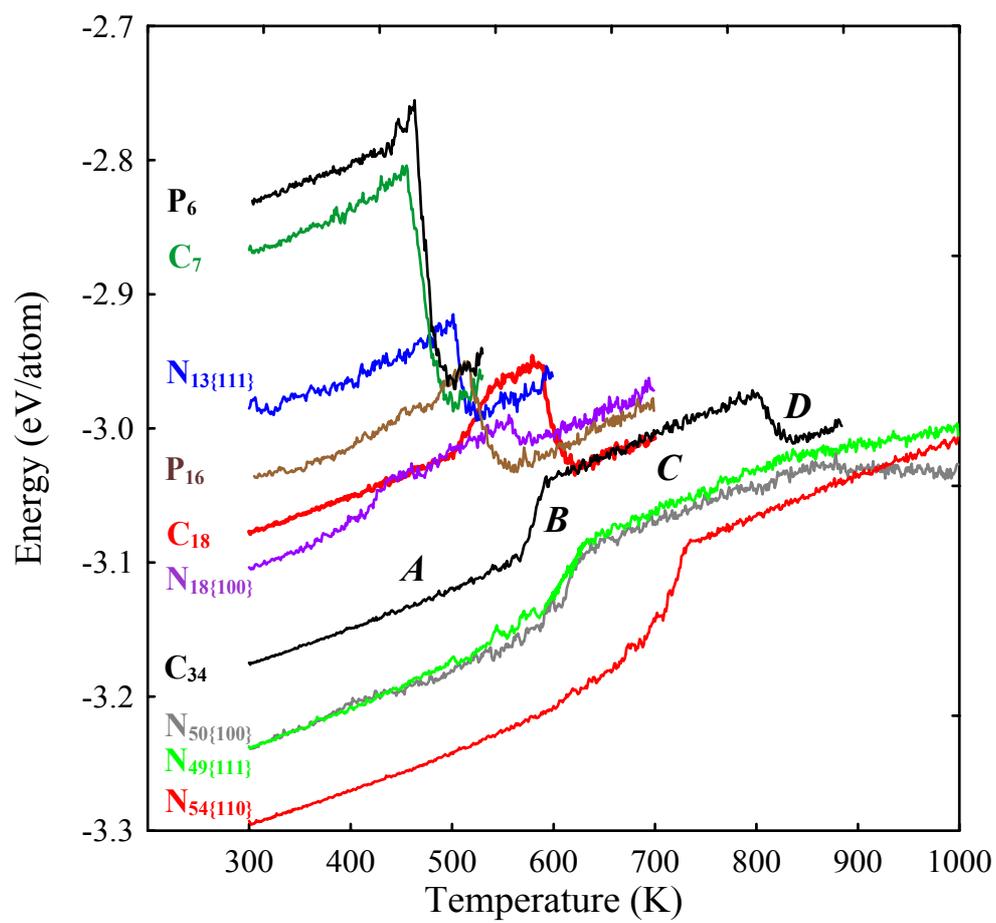

Figure 1. (color)

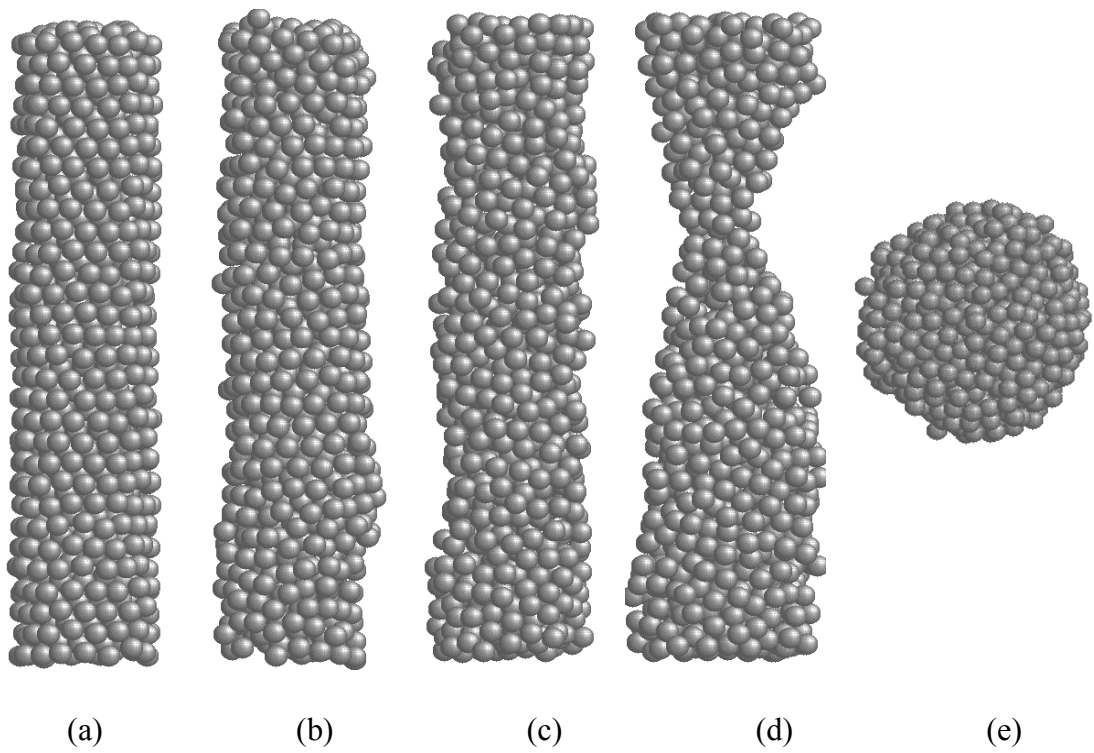

(a) (b) (c) (d) (e)

Figure 2.

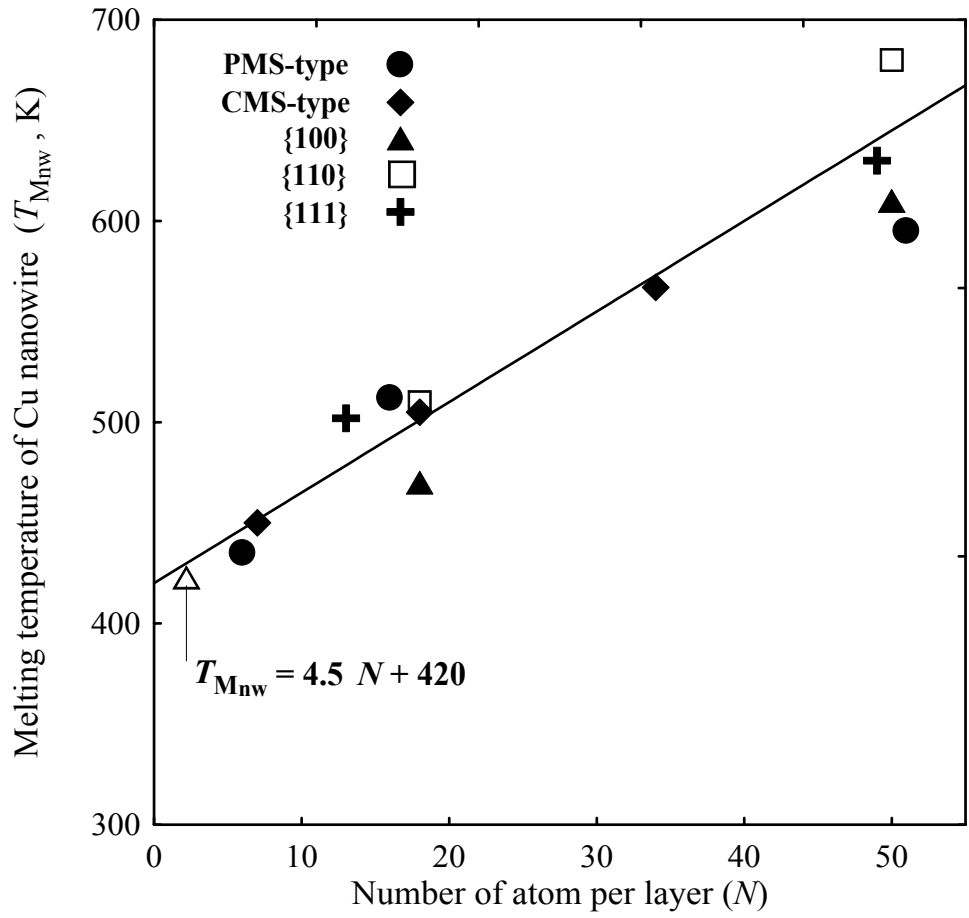

Figure 3.